\documentclass{article}

\usepackage{fancyhdr}
\usepackage{arxiv}

\usepackage[utf8]{inputenc} % allow utf-8 input
\usepackage[T1]{fontenc}    % use 8-bit T1 fonts
\usepackage{hyperref}       % hyperlinks
\usepackage{url}            % simple URL typesetting
\usepackage{booktabs}       % professional-quality tables
\usepackage{amsfonts}       % blackboard math symbols
\usepackage{amsmath}        % for \text in math mode
\usepackage{nicefrac}       % compact symbols for 1/2, etc.
\usepackage{algorithm}

\usepackage{algpseudocode} % for \While, \EndWhile, etc.
\usepackage{microtype}      % microtypography
\usepackage{lipsum}		% Can be removed after putting your text content
\usepackage{float} % for [H] float placement
\usepackage{graphicx}
\usepackage{natbib}
\usepackage{doi}

\title{Average Distance Approximation for Static Large Graphs}

\author{ {Kartikey Ahlawat}\\
	Leiden Institute of Advanced Computer Science \\
	Leiden University\\
	\texttt{kartikeyprit@gmail.com} \\
}	

\providecommand{\undertitle}{}

\hypersetup{
  pdftitle={Average Distance Approximation for Static Large Graphs},
  pdfsubject={average distance, mean distance, shortest path, graph sampling, node selection, social network analysis, network science},
  pdfauthor={Kartikey Ahlawat},
  pdfkeywords={average distance, mean distance, shortest path, graph sampling, node selection, social network analysis, network science},
}

\begin{document}
\maketitle

\begin{abstract}
	Calculating average distances in large-scale networks is computationally intensive and constrained by limited main memory, posing a significant challenge in graph analytics. This study explores and evaluates two primary approaches for estimating average distances: a graph sampling-based method (Random Walk) and landmark-based methods, including the Size Estimation Framework (SEF) and the Eppstein-Wang (EW) algorithm.\\
Random Walk was found to be unreliable for small sample sizes and computationally expensive for larger ones, requiring at least 15\% of nodes for accuracy.\\
Landmark-based approaches, leveraging probabilistic data structures like HyperLogLog for memory-efficient neighbor exploration, demonstrated superior performance. Among these, the SEF algorithm offers better memory efficiency, while the EW algorithm achieves higher accuracy with lower computation time.\\ 
Experiments on static, undirected, and unweighted graphs (both unipartite and bipartite) revealed that the EW algorithm produced results with an error margin as low as 0.02\%. Additionally, a subset of 100 randomly selected nodes was sufficient for accurate estimations in most large graphs.\\
The findings indicate that the EW algorithm provides a practical and scalable solution for average distance estimation, with higher reliability on unipartite graphs compared to bipartite graphs.
\end{abstract}

% keywords can be removed

\section{Introduction}
When dealing with graph statistics computing average/mean distance is arguably the most time and memory-consuming operation. When computing it for a large graph (having hundreds of thousands of nodes) it becomes almost impossible (time complexity of O($n^{\text{2}}$)).\\ 
But calculating this metric opens up significant opportunities for understanding various important network properties. Let's explore the significance of determining the average distance in graph theory. The average distance of a graph offers valuable insights into its fundamental characteristics, including:
\begin{itemize}
    \item How well connected is the graph: A smaller average distance implies that the nodes are well-connected and close to each other, demonstrating small-world properties, whereas a larger value indicates that nodes are dispersed. 
    \item Plotting distance distribution: The average distance value directly cannot tell about the distribution, but the method of plotting distance distribution is the same as calculating average distance (finding the shortest path between all nodes). Distance distribution can highlight isolated clusters or components of the graph if any. Also, by looking at the plot we can get the idea about graph uniformity.
    \item Providing a better understanding of various network properties across a range of domains, such as graph structure identification, connectivity assessment, network efficiency analysis, and social network evaluation.
    \item Robustness of network: generally graphs with low average distance are considered to be more robust and resilient to any kind of disruption as nodes are well connected and if any link is broken or corrupted there are other channels (edges) to substitute.
\end{itemize}

The graph's distance from node A to B refers to the minimum number of edges that must be traversed to reach node B. This could also be called the shortest path distance from A to B. The average distance can be called the average of all shortest path distances between all possible pairs of nodes. The mathematical notation can be seen in Equation~\ref{eq:avg_distance}. \\

Currently, there are two popular approaches to compute the average distance:
\begin{itemize}
    \item \textbf{Graph Sampling:} This approach suggests sampling smaller graphs from the original using sampling algorithms like Random Walk, Random Node Sampling, etc. And compute the average distance for the sampled graph.
    \item \textbf{Node Subset:} Here, we compute the shortest path from the subset of nodes to all the nodes of the original graph.
\end{itemize}
Both approaches can be performed by 'NetworkX' and 'igraph', an open-source Python library used for studying graph networks. They have a built-in function that computes the average distance utilizing Breadth First Search (BFS) under the hood to find the shortest path.\\
More on each approach is discussed in the section: 'Related Work'.
\\
This paper addresses the issue of computation time when computing the average distance for a large static graph. The paper explores different approaches to computing average distance with comparison. In the end, we conclude with the best approach as per the evidence and observation acquired during experiments.\\
Will focus on Static-Undirected-Unweighted graphs (both Unipartite \& Bipartite), as these are the most common type out there. Breaking down each component for better understanding:
\begin{itemize}
    \item \textbf{Static:} property of graph that goes on to show that the graph does not change over time. It has remained the same (number of node \& edges) from the time of its creation.
    \item \textbf{Undirected:} property of graph that goes on to show that edges are not directed, if there exists a path from $A \to B$ then the same path can be taken from $B \to A$. 
    \item \textbf{Unweighted:} Graph with edges holding no weight. All the edges hold an equal traversal cost of 1 (by default).  
    \item \textbf{Unipartite:} It represents a normal graph where there are no restrictions on node linkage. Some examples are social media and transportation networks.
    \item \textbf{Bipartite:} In a bipartite graph, the nodes from one disjoint set connect to the nodes of another disjoint set but not with each other from the same set. One of its examples could be the user-item relationship between Netflix's viewer-movie and Amazon's user-product.
\end{itemize}

\section{Related work}
The field of computing estimation for resource-intensive graph properties for static undirected large graphs has been extensively studied. Various approaches were suggested that balance the accuracy with computation time. Some of the approaches were not introduced to compute the average distance of the graph. Still they were found to be useful for our case. Here, we will review some of the main approaches discussed in the literature along with their strengths and limitations.
\begin{itemize}
    \item \textbf{Naive approach:} The direct calculation of average distance can be achieved using Breadth-First Search (BFS) or by utilizing NetworkX's predefined function, such as 'average \_shortest\_path\_length', which internally employs BFS. This approach averages the shortest path distances between all the node pairs of the original large graph.
    \begin{itemize}
        \item \textbf{Strength:} Gives exact average distance value. 
        \item \textbf{Limitation:} Computationally expensive with quadratic time complexity (O($n^{\text{2}}$), where n is the number of nodes in a large graph). This could only be considered if the graph is very small, but for real-world graphs (large) this approach is not suited. Moreover, if the graph is disconnected then using the NetworkX function will through an error and manual BFS will give incorrect results.
    \end{itemize}
    \item \textbf{Largest Connected Component approach:} Here, from the original large graph, we extract the largest connected component and calculate the average distance using Breadth First Search (BFS) or predefined function (average\_shortest\_path \_length). This approach averages the shortest path distances between all the node pairs of the sampled largest connected component.
    \begin{itemize}
        \item \textbf{Strength:} If a graph has disconnected components it is better to compute the average distance on its largest connected component instead of the whole graph.
        \item \textbf{Limitation:} This approach will also be Computationally expensive, as the large connected component of a large graph is also considerably large for calculating the shortest path among all its pairs. It could only be considered if the graph is very small.
    \end{itemize}
    \item \textbf{Graph Sampling:} This is also called sampled based approach. It allows us to derive a sample from the large graph that is a good representation of the original graph and holds its structural properties. Then, will calculate the shortest path distances between all the node pairs of the sampled graph. Which algorithm is best has remained a topic of argument, but in recent research 'Random Walk' was found to be best for sampling from static large graphs \cite{LargeGraphSampling}.
    \begin{itemize}
        \item \textbf{Strength:} Will take less time, as the number of nodes decreased for which the shortest path was to be calculated.
        \item \textbf{Limitation:} Sampled graph might be useful for taking approximations for parameters like clustering coefficient, degree distribution, etc. But for average distance which relies heavily on graph integrity, the sampled graph becomes unreliable for the estimation. Moreover, the implementation of graph sampling is complex and adds to the time complexity. 
    \end{itemize}
    \item \textbf{Node Subset:} This is also called landmark-based approach. It allows us to randomly select nodes from the large graph and then, calculate shortest path distances from nodes present in the subset to all the nodes of the large graph. This method is by far considered the best for estimating average distance by previous research \cite{inproceedings1}    
    \begin{itemize}
        \item \textbf{Strength:} More accurately captures the graph's overall structure and distances.
        \item \textbf{Limitation:} Random selection may result in sub-optimal node subset, that might not capture the global graph properties.
    \end{itemize}
    Algorithms for computing average distance using node subset:
    \begin{itemize}
        \item \textbf{Eppstein-Wang (EW) Algorithm:} take random node subset, apply BFS, and aggregate the distances to get the average.\cite{eppstein2000fastapproximationcentrality}
        \item \textbf{Size Estimation Framework (SEF):} This approach was introduced by Edith Cohen in his research paper \cite{COHEN1997441}. It was specifically applied for finding transitive closures but can also be applied to finding average distance as done in \cite{10.1007/978-3-642-19754-3_11}
    \end{itemize}
\end{itemize}
This paper will mainly experiment with the landmark-based approaches as they tend to yield better accuracy in comparison to sampled-based. Moreover, it retains overall distance property better. Still, results from the sampled-based approach will be included to see how much better landmark-based approach is. Furthermore, experiments will be done to see which node selection method provides the best accuracy in arguably less time. 

% briefly discuss other papers related to this work, or previous work describing other approaches for the same problem. End with a statement on how your paper contributes to these works.

\section{Preliminaries}
\begin{enumerate}

\item Given a graph G = (V, E), where V is the set of nodes and E is the set of edges, let $d(u,v)$ denote the shortest path between the two nodes $u$ and $v$. The average distance $\bar{d}$ over all pairs of nodes is computed by:
    \begin{equation}
    \bar{d} = \frac{1}{n \cdot (n-1)} \sum_{\substack{u,v \in V \\ u \neq v }} d(u,v)
    \label{eq:avg_distance}
    \end{equation}
    Where:
    \begin{itemize}
        \item n is the number of nodes.
        \item \( \sum_{\substack{u,v \in V \\ u \neq v }} \) sums the shortest path for all pairs of the nodes (u, v).
        \item This sum gets divided by the count of the total number of pairs, excluding self-pairing (v,v).
    \end{itemize}
\item Number of nodes in a sampled graph or node subset is represented by \(n'\).
\item Sampled graph is represented by $G_\text{\(n'\)}$, where \(n'\)<< n.
\item LCC stands for Largest Connected Component.
\end{enumerate}
% Necessary notation, formal definitions, if needed theorems, and a problem statement. Explain nontrivial notation upon first use. If applicable, give a formal problem statement and say something about space and time complexity. 

\section{Approach}
Implement SEF and EW Algorithm over 4 Unipartite-Static-Undirected-Unweighted Graphs and 4 Bipartite-Static-Undirected-Unweighted Graphs. Each graph will experiment with node subset sizes of 1, 5, 10, 30, 40, 100, and 1000. Furthermore, for each node subset size, the respective algorithm is tested four times to check consistency. To get a more reliable result the average over the four iterations will be taken to represent the result (average distance). The algorithm's performance will be evaluated based on node subset size, computation time, average distance estimation, standard deviation (over 4 iterations), and error\% to actual average distance.
We will also check the performance of the graph sampling, but will not go into deep experiments as in previous literature studies it has already been established that this is not an efficient and correct approach for average distance estimations as it can not correctly capture the graph distance properly \cite{LargeGraphSampling}. Will only see how bad it performs in comparison to the landmark-based approaches like SEF and EW algo.

% Factually present your solution to the problem studied in the paper. This may include a repetition in your own words of the original paper that you studied. Strive to have at least one explanatory picture that explains the approach, perhaps using the tikz package, or more algorithmically using the algorithm(ic) package. Distinguish between previously introduced methods and your contribution clearly, for example by using subsections named after the various (components of the) approach(es).  
			
\section{Data}
We selected open-source graph data from Konect. It also has graph statistics along with each graph like volume, node \& edge count, average degree, diameter, and most importantly average distance. During data searching our focus was on finding graphs that have average distance already reported under the statistics, so this way we already have the baseline and do not need to compute the actual value before jumping to experimentation approaches. Having the actual average distance for all our graphs saved us a huge amount of time and computational resources. Additionally, all the selected graphs have passed the consistency check. Moreover, no loops and multiple edges are present in the chosen graphs. Having them causes overhead computation and the algorithm might give an estimation with high absolute error if it is not handled explicitly.
\\
In total we have shortlisted 8 static-undirected-unweighted graphs among these half are unipartite and the other half are bipartite.\\

Here are some details about the Unipartite-Static-Undirected-Unweighted Graphs we used:
\begin{enumerate}
    \item \textbf{Flixster:} It is a movie rating website, where apart from rating movies users can connect with others who have similar taste in movies as you. In its graph network, nodes represent the users, and the edge represents friendship. Neither loop nor multiple edges are present \cite{konect_flixster}.
    
    \item \textbf{Skitter:} This was the initiative taken by CAIDA (Center for Applied Internet Data Analysis) to monitor and understand autonomous system's network topology. The nodes represent autonomous systems and the edge is the connection between them. Neither loop nor multiple edges are present \cite{konect_as_skitter}.
    
    \item \textbf{Youtube:} This represents the network between YouTube users. The nodes represent users and the edge is the connection between them. In YouTube, if you subscribe to any channel or user, then there is a connection established between you two. Neither loop nor multiple edges are present \cite{konect_youtube_growth}.

    \item \textbf{Orkut:} This was a social media website started by Google. The nodes represent users and the edge is the link between them. Neither loop nor multiple edges are present \cite{konect_orkut_links}.
\end{enumerate}
Table~\ref{table:Overview_Unipartite_Stats} illustrates the stats for the Unipartite datasets used for the experimentation.\\

\begin{table*}[ht!]
    \centering
    \caption{Overview of Unipartite Datasets Stats}
    \begin{tabular}{|l|c|c|c|c|}
        \hline
        \textbf{Parameters} & \textbf{Flixster Stats}  & \textbf{Skitter Stats} & \textbf{Youtube Stats} & \textbf{Orkut Stats} \\ \hline
        Node count (n) & 2,523,386 & 1,696,415 & 3,223,589 & 3,072,441 \\ \hline
        Edge Count & 7,918,801 & 11,095,298 & 9,375,374 & 117,184,899 \\ \hline
        Loop Count & 0 & 0 & 0 & 0 \\ \hline
        Triangle Count & 7,897,122 & 28,769,868 & 12,226,580 & 627,577,371 \\ \hline
        Square Count & 454,368,595 & 62,769,198,018 & 9,890,851,109 & Not Available \\ \hline
        Diameter & 8 & 31 & 31 & 10 \\ \hline
        Average Degree & 6.27 & 13.08 & 5.81 & 76.28 \\ \hline
        Size of LCC & 2,523,386 & 1,694,616 & 3,216,075 & 3,072,441 \\ \hline
        Clustering Coef. & 0.013 & 0.005 & 0.001 & 0.041 \\ \hline
        Power Law Exponent & 2.59 & 1.61 & 2.33 & 1.27 \\ \hline
        Average Distance & 4.815 & 5.036 & 5.291 & 4.266 \\ \hline
    \end{tabular}
    \label{table:Overview_Unipartite_Stats}
\end{table*}

Here are the details about the Bipartite-Static-Undirected-Unweighted Graphs we used:
\begin{enumerate}
    \item \textbf{Flickr:} It is a photo and video-sharing website. In its graph network nodes represent users and groups and edge as membership. It is bipartite because no user-user connection only user-group connections. No multiple edges are present \cite{konect_flickr_groupmemberships}.

    \item \textbf{DBLP:} It is a computer science bibliography, it stands for Database Systems and Logic Programming. The nodes represent authors and publications, and the edge connects the author to one of its publications. No multiple edges are present \cite{konect_dblp_author}.

    \item \textbf{CiteSeer:} It is an authorship network. The nodes represent authors and publications, and the edge connects an author to their publication \cite{konect_citeseer}.

    \item \textbf{Occupations from DBpedia:} DBpedia extracts information from Wikipedia and shows it graphically. The nodes represent people and occupations, and the edge connects a person to their respective occupation. No multiple edges are present \cite{konect_dbpedia_occupation}.
\end{enumerate}

Table~\ref{table:Overview_Bipartite_Stats} illustrates the stats for the Bipartite datasets used for the experimentation.\\

\begin{table*}[ht!]
    \centering
    \caption{Overview of Bipartite Datasets Stats}
    \resizebox{\textwidth}{!}{
    \begin{tabular}{|l|c|c|c|c|}
        \hline
        \textbf{Parameters} & \textbf{Flickr's Stats}  & \textbf{DBLP Stats} & \textbf{CiteSeer Stats} & \textbf{Occupations from DBpedia Stats} \\ \hline
        Node count (n) & 499,610 & 7,577,304 & 286,748 & 229,307 \\ \hline
        Bipartite Group A & (User) 395,979 & (Author) 1,953,085 & (Author) 105,353 & (People) 127,577 \\ \hline
        Bipartite Group b & (Group) 103,631 & (Publication) 5,624,219 & (Publication) 181,395 & (Occupation) 101,730 \\ \hline
        Edge Count & 8,545,307 & 12,282,059 & 512,267 & 250,945 \\ \hline
        Square Count & 35,258,013,095 & 31,673,959 & 1,005,817 & 24,509,245 \\ \hline
        Diameter & 18 & 48 & 34 & 24 \\ \hline
        Average Degree & 34.2 & 3.24 & 3.57 & 2.18 \\ \hline
        Average Degree of Group A subset & 21.5 & 6.28 & 4.86 & 1.96 \\ \hline
        Average Degree of Group B subset & 82.45 & 2.18 & 2.82 & 2.46 \\ \hline
        Size of LCC & 470,345 & 6,735,203 & 269,000 & 143,220 \\ \hline
        Power Law Exponent & 1.58 & 2.33 & 2.07 & 4.14 \\ \hline
        Average Distance & 4.43 & 10.97 & 10.25 & 5.19 \\ \hline
    \end{tabular}
    }
    \label{table:Overview_Bipartite_Stats}
\end{table*}

% what datasets did you use, what types of networks do they represent, where did you obtain the data? did you do any processing? give a table describing data characteristics, such as number of nodes, edges, average degree, etc. 
% Is the data possible biased and how does this affect the experiments? 

\section{Experiments}
We started our experimentation with a graph sampling approach by applying the Random Walk method to the Flixster, Skitter, and YouTube datasets from the unipartite graph category (see Table~\ref{table:Overview_Unipartite_Stats}) and subsequently calculating shortest paths using parallel processing. In the Random Walk process, we start from a randomly selected node and traverse its neighbors step by step. If a node has no neighbors left to explore, we select a new random starting node and continue the traversal. This process is repeated until the desired sample size is achieved.\\
The performance was evaluated based on node subset size, computation time, average distance estimation, standard deviation (over 3 iterations), and error\% to actual average distance.\\
The Flixster, Skitter, and YouTube datasets from the unipartite graph category are significantly large. Hence, we limited our analysis to sample sizes of 0.1\% and 0.2\%, as larger sample sizes required substantial computational resources and excessive execution time.\\
Refer to Algorithm~\ref{algo: Graph Sampling and Distance Calculation}, while the corresponding results are presented in Table~\ref{table: Statistics for Graph Sampling}.

\begin{algorithm} [ht]
    \caption{Graph Sampling and Distance Calculation}
    \label{algo: Graph Sampling and Distance Calculation}
    \begin{algorithmic}[1]
    \Require Graph $G$, sample fraction $f$, number of workers $num\_workers$
    \Ensure Estimated mean shortest path distance
    
    \State $total\_nodes \gets G.vcount()$
    \State $sample\_size \gets \max(1, \lfloor f \cdot total\_nodes \rfloor)$
    
    \State \Comment{Step 1: Perform Random Walk Sampling in Parallel}
    \State $nodes\_per\_worker \gets \lfloor sample\_size / num\_workers \rfloor$
    \State $start\_nodes \gets$ Randomly sample $num\_workers$ nodes from $G$
    \State $tasks \gets [(G, start\_node, nodes\_per\_worker) \; \forall \; start\_node \in start\_nodes]$
    \State $sampled\_node\_ids \gets$ Execute $random\_walk\_sample$ for all $tasks$ in parallel
    \While{$|sampled\_node\_ids| < sample\_size$}
        \State $additional\_nodes \gets random\_walk\_sample(G, $ \\ \hspace{1.5em}$\text{random node from } G, sample\_size - |sampled\_node\_ids|)$
        \State $sampled\_node\_ids \gets sampled\_node\_ids \cup additional\_nodes$
    \EndWhile
    
    \State \Comment{Step 2: Create a Subgraph of Sampled Nodes}
    \State $sampled\_graph \gets G.subgraph(sampled\_node\_ids)$
    \State $mapped\_nodes \gets \{0, 1, ..., sampled\_graph.vcount() - 1\}$
    
    \State \Comment{Step 3: Calculate Shortest Paths in Parallel}
    \State $node\_pairs \gets$ All unique pairs of $mapped\_nodes$
    \State $chunk\_size \gets \lfloor |node\_pairs| / num\_workers \rfloor$
    \State $tasks \gets$ Divide $node\_pairs$ into chunks of size $chunk\_size$
    \State $shortest\_paths \gets$ Execute $calculate\_shortest\_path$ for all $tasks$ in parallel
    
    \State \Comment{Step 4: Calculate Mean Shortest Path Distance}
    \State $path\_lengths \gets \text{Values of } shortest\_paths \text{ excluding } \infty$
    \If{$|path\_lengths| > 0$}
        \State $mean\_distance \gets \frac{\sum(path\_lengths)}{|path\_lengths|}$
    \EndIf
    
    \State \Return $mean\_distance$
    
    \end{algorithmic}
\end{algorithm}

\begin{table*}[ht!]
    \centering
    \caption{Statistics for Graph Sampling}
    \resizebox{\textwidth}{!}{
    \begin{tabular}{|c|c|c|c|c|c|c|c|c|}
    \hline
        \textbf{Dataset} & \textbf{Node Subset Size} & \textbf{\% of total}	& \textbf{Computation Time} &	\textbf{Av. Distance (mean over 3 iter.)} & \textbf{\( \sigma \)} & \textbf{Error}	& \textbf{Error\% w.r.t. actual} \\ \hline
        Flixster &	2523 &	0.1\% & 3.5 min & 5.43 & 0.22 &	0.62&	12.84 \\ \hline
        Flixster &	5046 &	0.2\% & 30 min & 4.76 & 0.09 & 0.05 & 1.07 \\ \hline
        Skitter &	1629 &	0.1\% & 1 min & 5.09 &	0.25 &	0.054 &	1.07 \\ \hline
        Skitter &	3392 &	0.2\% & 10 min & 4.65 &	0.17 &	0.38 & 7.59 \\ \hline
        Youtube &	3223 &	0.1\% & 9 min & 3.46 & 0.36 & 1.83 & 34.67 \\ \hline
        Youtube &	6447 &	0.2\% & 99 min & 3.52 & 0.37 & 1.77 & 33.41 \\ \hline
    \end{tabular}
    }
    \label{table: Statistics for Graph Sampling}
\end{table*}
We opted not to proceed with this method due to its unreliability with very small sample sizes. When the sample size is small, the results vary significantly depending on the percentage of the sample size relative to the original dataset and the number of nodes selected during the random walk. Our experiments revealed that as the dataset size increased, the results became increasingly unreliable. Specifically, the error margins grew substantially when using the same 0.1\% and 0.2\% sample sizes on larger datasets. Reliable results generally require larger sample sizes (at least 15\%), as recommended in the study~\cite{LargeGraphSampling}.  However, using such larger samples introduces significant computational overhead, making the problem computationally too expensive.\\

Subsequently, we began exploring landmark-based approaches such as SEF and EW Algo. SEF use the concept of exploring neighbors within n-hops of their reach with the help of HyperLogLog. HyperLogLog is a probabilistic data structure that uses hashing to store the elements, used to estimate the cardinality in large graphs for being memory efficient as it consumes constant memory irrespective of the graph size.
We can use SEF indirectly to compute the average distance by leveraging the node count at each hop. By default, it returns the list of unique node counts that can be reached under n-hops from the randomly sampled node or subset of nodes. If max\_hop is set to 5 it will return a list of size 5. \\

Our proposal:\\
SEF output (S) w.r.t. v, where v is a randomly selected node from a large graph = [100, 300, 600, 1000, 1500]\\
Reachable nodes in 1-hop from v = 100\\
Reachable nodes in 2-hops from v = 300-100 = 200\\
Reachable nodes in 3-hops from v = 600-300 = 300\\
...\\
Nodes at distance n from v = S$_{n}$ - S$_{n-1}$\\

The average distance for 1 sampled node becomes:\\
\[
\text{Average Distance} = \frac{\sum_{i=1}^{\text{max\_hops}} \text{nodes\_at\_distance}[i] \times i}{\sum_{i=1}^{\text{max\_hops}} \text{nodes\_at\_distance}[i]}
\]\\

The average distance for n sampled node becomes:\\
\[
\text{Average Distance} = \frac{\sum_{j=1}^{\text{n}} \sum_{i=1}^{\text{max\_hops}} \text{nodes\_at\_distance}[j, i] \times i}{\sum_{j=1}^{\text{n}} \sum_{i=1}^{\text{max\_hops}} \text{nodes\_at\_distance}[j, i]}
\]

Refer Algorithm~\ref{algo: SEF Algo}

\begin{algorithm}[H]
\caption{Estimate Average Distance using Size Estimation Framework (SEF)}
\label{algo: SEF Algo}
\begin{algorithmic}[2]
\Require Graph $G$ with its largest connected component $largest\_cc$, max hops $max\_hops$, sample size $sample\_size$
\Ensure Estimated average distance

\State $nodes \gets$ randomly sample $sample\_size$ nodes from $largest\_cc$
\State $cumulative\_counts \gets []$

\For{$hop \gets 1$ to $max\_hops$}
    \State $hll \gets$ Initialize HyperLogLog with error rate $0.01$
    \For{$node \in nodes$}
        \State $reachable\_nodes \gets$ Neighborhood of $node$ in $largest\_cc$ within distance $hop$
        \For{$reachable\_node \in reachable\_nodes$}
            \State $hll.add(\text{string}(reachable\_node))$
        \EndFor
    \EndFor
    \State $cumulative\_counts.append(len(hll))$
\EndFor

\State $exact\_counts \gets [cumulative\_counts[0]]$
\For{$i \gets 1$ to $|cumulative\_counts| - 1$}
    \State $exact\_counts.append(cumulative\_counts[i] - cumulative\_counts[i - 1])$
\EndFor

\State $weighted\_sum \gets \sum_{i=1}^{|exact\_counts|} i \times exact\_counts[i]$
\State $total\_nodes \gets \sum_{i=1}^{|exact\_counts|} exact\_counts[i]$

\If{$total\_nodes > 0$}
    \State \Return $weighted\_sum / total\_nodes$
\Else
    \State \Return $0$
\EndIf

\end{algorithmic}
\end{algorithm}

Moving on to Eppstein-Wang Algorithm, which is pretty straightforward in comparison to SEF. A subset of nodes is randomly selected and then the shortest path length is calculated from each of them to all the nodes of the large graph. Finally, the average of the sum gives us the estimation of the average distance for the large graph. Refer Algorithm~\ref{algo: EW Algo}

\begin{algorithm}[H]
\caption{Estimate Average Distance using Eppstein-Wang Algorithm}
\label{algo: EW Algo}
\begin{algorithmic}[3]
\Require Graph $G$ with its largest connected component $largest\_cc$
\Ensure Average distance of all nodes in $largest\_cc$

\Function{ComputeShortestPaths}{node}
    \State \Return $largest\_cc.shortest\_paths(node)[0]$
\EndFunction

\State $all\_distances \gets$ \Call{Map}{\textsc{ComputeShortestPaths}, all nodes in $largest\_cc$}

\State $distances \gets []$
\For{each $sublist \in all\_distances$}
    \For{each $dist \in sublist$}
        \If{$dist > 0$} \Comment{Ignore self-loops}
            \State $distances.append(dist)$
        \EndIf
    \EndFor
\EndFor

\State $average\_distance \gets \frac{\sum_{dist \in distances} dist}{|distances|}$

\State \Return $average\_distance$
\end{algorithmic}
\end{algorithm}

Additionally, during experimentation on algorithms using different sample sizes we realized that NetworkX is not suitable for large graphs. Instead, igraph is a better option it increases speed (written in C) and is space efficient (compact internal data representation). So, all the experimentation was implemented using igraph.\\

\subsection{SEF vs EW Algorithm}
We applied the SEF and EW algorithms on YouTube graph data to get the estimated average distance. For each node subset size the algorithm was run 4 times to ensure consistency. Mean of 'average distance' was taken over the 4 iterations and calculated error\% to the actual 'average distance' value.
\[\text{Error\%} = \left(\frac{\lvert \text{actual\_average\_dist} - \text{mean\_average\_dist} \rvert}{\text{actual\_average\_dist}}\right) \times 100
\]
Please refer to Table-\ref{table: SEF Test Run Result} and Table-\ref{table: EW Algorithm Test Run Result} respectively, for the results.

\begin{table}[ht!]
        \centering
        \caption{SEF Test Run Result on YouTube Data}
        \begin{tabular}{|c|c|c|}
        \hline
            \textbf{Node Subset Size} & \textbf{Computation Time(min)} & \textbf{Error\%} \\ \hline
           	1& 2&	5.74 \\ \hline
           5&	 3&	3.75 \\ \hline
           10&	 5&	2.19 \\ \hline
           30&	 20&	18.02 \\ \hline
        \end{tabular}
        \label{table: SEF Test Run Result}
    \end{table}

\begin{table}[ht!]
        \centering
        \caption{EW Algo Test Run Result on YouTube Data}
        \begin{tabular}{|c|c|c|}
        \hline
            \textbf{Node Subset Size} & \textbf{Computation Time(sec)} & \textbf{Error\%} \\ \hline
           	1& 1&	4.93 \\ \hline
           5&	 3&	5.16 \\ \hline
           10&	 6&	9.56 \\ \hline
           30&	 18&	4.46 \\ \hline
        \end{tabular}
        \label{table: EW Algorithm Test Run Result}
    \end{table}

Observing both tables (\ref{table: SEF Test Run Result} \& \ref{table: EW Algorithm Test Run Result}) it is clear that SEF is more time-consuming, taking minutes for the same node subset size for which EW gives output within a few seconds.\\
The SEF algorithm takes more time (even if the diameter is known, i.e., max\_hop) than the EW algorithm, but is space-optimized. So if memory efficiency is more priority than time then use SEF otherwise EW algorithm. Interestingly, with the increase in node subset size there was no linear decline observed in the error\%. For node subset sizes 5 and 10, SEF performed better with lower error\%, but overall EW algorithm attained lower error\%.

\subsection{EW Algorithm on Unipartite Graph}
As mentioned in the 'Approach' and 'Data' sections of this paper, we have 8 static-undirected-unweighted graphs among these half are unipartite and the other half are bipartite (4-4 each). In this subsection will be applying EW algorithm over the unipartite graphs to see how it performs.
\begin{itemize}
    \item \textbf{Youtube Graph Data:} As observed from the table-\ref{table: Statistics on YouTube Data}, with an increase in the node subset size, the standard deviation decreases, pointing out that for all 4 iterations, the average distance values are considerably close to each other showing algorithm's high confidence. However, Error\% does not show an inverse relationship with the subset size, remaining similar as the subset size increases. Minimum error\% of 4.41 observed at node subset size of 100.
    \item \textbf{Flixster Graph Data:} As observed from the table-\ref{table: Statistics on Flixster Data}, standard deviation shows a negative correlation with subset size but it is not as strong as it was in YouTube data. However, Error\% shows an inverse relationship with the subset size up to size 10, after that it increases with the subset size. Minimum error\% of 1.24 observed at node subset size of 10.
    \item \textbf{Skitter Graph Data:} As observed from the table-\ref{table: Statistics on Skitter Data}, standard deviation shows a strong negative correlation with subset size. However, Error\% does not show any inverse relationship with the subset size. Minimum error\% of 0.02 observed at node subset size of 40.
    \item \textbf{Orkut Graph Data:} As observed from the table-\ref{table: Statistics on Orkut Data}, standard deviation shows a strong negative correlation with subset size. However, Error\% does not show any inverse relationship with the subset size. Minimum error\% of 0.31 observed at node subset size of just 1.
\end{itemize}

\begin{table*}[htbp]
    \centering
    \caption{Statistics on YouTube Data}
    \resizebox{\textwidth}{!}{
    \begin{tabular}{|c|c|c|c|c|c|c|c|}
    \hline
        \textbf{Node Subset Size} & \textbf{\% of total}	& \textbf{Computation Time} &	\textbf{Av. Distance (mean over 4 iter.)} & \textbf{\( \sigma \)} & \textbf{Error}	& \textbf{Error\% w.r.t. actual} \\ \hline
        1&	3e-5\%& 1 sec&	 5.03&	1.15&	0.26&	4.93 \\ \hline
        5&	1e-4\%& 3 sec&	 5.01&	0.24&	0.27&	5.16 \\ \hline
        10&	3e-4\%& 6 sec&	 4.78&	0.19&	0.5&	9.56 \\ \hline
        30&	9e-4\%& 18 sec&	 5.05&	0.14&	0.23&	4.46 \\ \hline
        40&	1e-3\%& 24 sec&	 4.98&	0.13&	0.3& 5.83 \\ \hline
        100&	3e-3\%& 1 min&	 5.05&	0.04&	0.23&	4.41 \\ \hline
        1000&	3e-2\%& 7 min&	 4.98&	0.008&	0.3&	5.78 \\ \hline
    \end{tabular}
    }
    \label{table: Statistics on YouTube Data}
\end{table*}

\begin{table*}[htbp]
    \centering
    \caption{Statistics on Flixster Data}
    \resizebox{\textwidth}{!}{
    \begin{tabular}{|c|c|c|c|c|c|c|c|}
    \hline
        \textbf{Node Subset Size} & \textbf{\% of total}	& \textbf{Computation Time} &	\textbf{Av. Distance (mean over 4 iter.)} & \textbf{\( \sigma \)} & \textbf{Error}	& \textbf{Error\% w.r.t. actual} \\ \hline
        1&	3e-5\%& 1 min&	 5.17&	0.18&	0.36&	7.52 \\ \hline
        5&	1e-4\%& 2 min&	 5.06&	0.11&	0.24&	5.14 \\ \hline
        10&	3e-4\%& 3 min&	 4.87&	0.12&	0.06&	1.24 \\ \hline
        30&	1e-3\%& 10 min&	 4.9&	0.09&	0.09&	1.92 \\ \hline
        40&	1e-3\%& 7 min&	 4.91&	0.16&	
        0.09& 2.02 \\ \hline
    \end{tabular}
    }
    \label{table: Statistics on Flixster Data}
\end{table*}

\begin{table*}[htbp]
    \centering
    \caption{Statistics on Skitter Data}
    \resizebox{\textwidth}{!}{
    \begin{tabular}{|c|c|c|c|c|c|c|c|}
    \hline
        \textbf{Node Subset Size} & \textbf{\% of total}	& \textbf{Computation Time} &	\textbf{Av. Distance (mean over 4 iter.)} & \textbf{\( \sigma \)} & \textbf{Error}	& \textbf{Error\% w.r.t. actual} \\ \hline
        1&	5e-5\%& 2 sec&	 5.07&	0.29&	0.03&	0.72 \\ \hline
        5&	2e-4\%& 3 sec&	 4.87&	0.22&	0.15&	3.14 \\ \hline
        10&	5e-4\%& 4 sec&	 5.11&	0.31&	0.08&	1.61 \\ \hline
        30&	1e-3\%& 8 sec&	 5.1&	0.18&	0.06&	1.27 \\ \hline
        40&	2e-3\%& 10 sec&	 5.03&	0.02&	0.001& 0.02 \\ \hline
        100&	5e-3\%& 22 sec&	 5&	0.04&	0.02&	0.56 \\ \hline
        1000&	5e-2\%& 3 min&	 5.09&	0.004&	0.05&	1.12 \\ \hline
    \end{tabular}
    }
    \label{table: Statistics on Skitter Data}
\end{table*}

\begin{table*}[htbp]
    \centering
    \caption{Statistics on Orkut Data}
    \resizebox{\textwidth}{!}{
    \begin{tabular}{|c|c|c|c|c|c|c|c|}
    \hline
        \textbf{Node Subset Size} & \textbf{\% of total}	& \textbf{Computation Time} &	\textbf{Av. Distance (mean over 4 iter.)} & \textbf{\( \sigma \)} & \textbf{Error}	& \textbf{Error\% w.r.t. actual} \\ \hline
        1&	3e-5\%& 8 min&	 4.25&	0.41&	0.01&	0.31 \\ \hline
        5&	1e-4\%& 30 min&	 4.06&	0.33&	0.2&	4.77 \\ \hline
        10&	3e-4\%& 2 min&	 4.25&	0.34&	0.01&	0.37 \\ \hline
        30&	9e-4\%& 12 min&	 4.1&	0.28&	0.16&	3.77 \\ \hline
        40&	1e-3\%& 22 min&	 4.12&	0.26&	0.14& 3.42 \\ \hline
        100&	3e-3\%& 2 hr&	 4.49&	0.27&	0.22&	5.25 \\ \hline
    \end{tabular}
    }
    \label{table: Statistics on Orkut Data}
\end{table*}

\begin{table*}[htbp]
    \centering
    \caption{Statistics on Flickr Data}
    \resizebox{\textwidth}{!}{
    \begin{tabular}{|c|c|c|c|c|c|c|c|}
    \hline
        \textbf{Node Subset Size} & \textbf{\% of total}	& \textbf{Computation Time} &	\textbf{Av. Distance (mean over 4 iter.)} & \textbf{\( \sigma \)} & \textbf{Error}	& \textbf{Error\% w.r.t. actual} \\ \hline
        1&	2e-4\%& 1 min&	 3.86&	0.28&	0.56&	12.81 \\ \hline
        5&	1e-3\%& 2 min&	 3.66&	0.6&	0.76&	17.21 \\ \hline
        10&	2e-3\%& 5 min&	 3.58&	0.39&	0.85&	19.18 \\ \hline
        30&	6e-3\%& 31 min&	 3.59&	0.48&	0.83&	18.9 \\ \hline
        40&	8e-3\%& 54 min&	 3.49&	0.35&	0.93& 21.1 \\ \hline
        100&	2e-2\%& 6 min&	 3.22&	0.3&	1.2&	27.14 \\ \hline
    \end{tabular}
    }
    \label{table: Statistics on Flickr Data}
\end{table*}

\begin{table*}[htbp]
    \centering
    \caption{Statistics on DBLP Data}
    \resizebox{\textwidth}{!}{
    \begin{tabular}{|c|c|c|c|c|c|c|c|}
    \hline
        \textbf{Node Subset Size} & \textbf{\% of total}	& \textbf{Computation Time} &	\textbf{Av. Distance (mean over 4 iter.)} & \textbf{\( \sigma \)} & \textbf{Error}	& \textbf{Error\% w.r.t. actual} \\ \hline
        1&	1e-5\%& 4 min&	 7.46&	0.36&	3.5&	31.92 \\ \hline
        5&	6e-5\%& 15 min&	 7.94&	0.56&	3.02&	27.59 \\ \hline
        10&	1e-4\%& 54 min&	 7.84&	0.47&	3.12&	28.48 \\ \hline
        30&	3e-4\%& 6 min&	 7.86&	0.33&	3.1&	28.28 \\ \hline
        40&	5e-4\%& 12 min&	 7.49&	0.35&	3.47& 31.65 \\ \hline
    \end{tabular}
    }
    \label{table: Statistics on DBLP Data}
\end{table*}

\begin{table*}[htbp]
    \centering
    \caption{Statistics on DBpedia Data}
    \resizebox{\textwidth}{!}{
    \begin{tabular}{|c|c|c|c|c|c|c|c|}
    \hline
        \textbf{Node Subset Size} & \textbf{\% of total}	& \textbf{Computation Time} &	\textbf{Av. Distance (mean over 4 iter.)} & \textbf{\( \sigma \)} & \textbf{Error}	& \textbf{Error\% w.r.t. actual} \\ \hline
        1&	4e-4\%& 20 sec&	 4.63&	0.73&	0.55&	10.69 \\ \hline
        5&	2e-3\%& 1 min&	 5.23&	1.29&	0.04&	0.91 \\ \hline
        10&	4e-3\%& 5 min&	 3.87&	0.29&	1.31&	25.38 \\ \hline
        30&	1e-2\%& 6 min&	 4.61&	1.54&	0.57&	11.07 \\ \hline
        40&	1e-2\%& 10 min&	 4.74&	0.68&	0.44& 8.62 \\ \hline
        100&	4e-2\%& 1 min&	 4.97&	1.57&	0.21& 4.19 \\ \hline
    \end{tabular}
    }
    \label{table: Statistics on DBpedia Data}
\end{table*}

\begin{table*}[htbp]
    \centering
    \caption{Statistics on CiteSeer Data}
    \resizebox{\textwidth}{!}{
    \begin{tabular}{|c|c|c|c|c|c|c|c|}
    \hline
        \textbf{Node Subset Size} & \textbf{\% of total}	& \textbf{Computation Time} &	\textbf{Av. Distance (mean over 4 iter.)} & \textbf{\( \sigma \)} & \textbf{Error}	& \textbf{Error\% w.r.t. actual} \\ \hline
        1&	3e-4\%& 20 sec&	 6.43&	0.21&	3.81&	37.19 \\ \hline
        5&	1e-3\%& 35 sec&	 6.73&	0.23&	3.51&	34.26 \\ \hline
        10&	3e-3\%& 1 min&	 6.52&	0.33&	3.72&	36.31 \\ \hline
        30&	1e-2\%& 5 min&	 6.47&	0.08&	3.78&	36.87 \\ \hline
        40&	1e-2\%& 8 min&	 6.08&	0.44&	4.17& 40.68 \\ \hline
        100&	3e-2\%& 1 min&	 6.64&	0.21&	3.6& 35.19 \\ \hline
    \end{tabular}
    }
    \label{table: Statistics on CiteSeer Data}
\end{table*}

\subsection{EW Algorithm on Bipartitie Graph}
In this subsection will be applying EW algo over the bipartite graphs to see how it performs.
\begin{itemize}
    \item \textbf{Flickr Graph Data:} As observed from the table-\ref{table: Statistics on Flickr Data}, the standard deviation does not show any correlation with subset size. Moreover, Error\% shows a proportionate relationship with the subset size. It is increasing with an increase of the subset size. Minimum error\% of 12.89 observed at node subset size of just 1.
    \item \textbf{DBLP Graph Data:} As observed from the table-\ref{table: Statistics on DBLP Data}, standard deviation and Error\% both show no correlation with subset size. Minimum error\% of 27.59 observed at node subset size of 5.
    \item \textbf{DBpedia Graph Data:} As observed from the table-\ref{table: Statistics on DBpedia Data}, standard deviation and Error\% show no correlation with subset size. Minimum error\% of 0.91 observed at node subset size of 5.
    \item \textbf{CiteSeer Graph Data:} As observed from the table-\ref{table: Statistics on CiteSeer Data}, standard deviation and Error\% show no correlation with subset size. Minimum error\% of 34.26 observed at node subset size of 5.
\end{itemize}
    
% Subsections on for example the experimental setup (which software, hardware and parameters did you choose), as well as the results of applying your approach to the data you described in preceding sections, leading to results that answer your research questions. You likely present some tables and figures. Remember captions and axis labels. 

% Think of incorporating the followng: What experiments can be used to compare, test and verify the suggested approaches? What do you measure in each experiment? Quality, running time, error size? Be precise. For datasets that perform either very well or not so well, try to find out why. Zoom in with additional experiments on noticeable results.  Can you relate the performance of the algorithms to some of the properties of the datasets? Or: can you define for which particular datasets a certain technique works well, and why? Make sure that the type of data/result that you want to communicate is suitable for the chosen presentation mode.

\section{Conclusion}
High computational time and limited main memory hinder the calculation of the average distance in large graphs. The issue has been addressed using landmark based-approaches, which estimate the average distance by calculating it for randomly selected subset of nodes. These methods have been applied to static, undirected, unweighted graphs (both Unipartite
\& Bipartite). \\
Eppstein Wang algorithm (EW Algorithm) was found to produce the most accurate result. Experiments showed that average distance estimations had an error as low as 0.02\%. Additionally, a node subset size of up to 100 was sufficient for most large graphs, as increasing the subset size generally did not improve accuracy. Overall, the algorithm provided better estimations for unipartite than bipartite graphs. \\
Future work could explore the performance of EW algorithm on large graphs beyond static, undirected, unweighted types. Furthermore, strategies to improve estimation accuracy for bipartite graphs requires more research.

\bibliographystyle{unsrtnat}
\bibliography{references}  

\end{document}